\documentclass[prb,preprint,showpacs,eqsecnum]{revtex4}
\usepackage{graphicx}
\usepackage{bm}
\begin{document}
\title{Dynamical and thermal effects in nanoparticle systems driven by a
rotating magnetic field}
\author{S.~I.~Denisov,$^{1,2}$ T.~V.~Lyutyy,$^{2}$ P.~H\"{a}nggi,$^{1}$
and K.~N.~Trohidou,$^{3}$}
\affiliation{$^{1}$Institut f\"{u}r Physik, Universit\"{a}t
Augsburg, Universit\"{a}tsstra{\ss}e 1, D-86135 Augsburg, Germany\\
$^{2}$Sumy State University, 2 Rimsky-Korsakov Street, 40007 Sumy, Ukraine\\
$^{3}$Institute of Materials Science, NCSR ``Demokritos,'' 15310 Athens,
Greece}


\begin{abstract}
We study dynamical and thermal effects that are induced in nanoparticle systems
by a rotating magnetic field. Using the deterministic Landau-Lifshitz equation
and appropriate rotating coordinate systems, we derive the equations that
characterize the steady-state precession of the nanoparticle magnetic moments
and study a stability criterion for this type of motion. On this basis, we
describe (i) the influence of the rotating field on the stability of the
small-angle precession, (ii) the dynamical magnetization of nanoparticle
systems, and (iii) the switching of the magnetic moments under the action of
the rotating field. Using the backward Fokker-Planck equation, which
corresponds to the stochastic Landau-Lifshitz equation, we develop a method for
calculating the mean residence times that the driven magnetic moments dwell in
the up and down states. Within this framework, the features of the induced
magnetization and magnetic relaxation are elucidated.
\end{abstract}
\maketitle

\section{INTRODUCTION}

The study of the dynamics of the nanoparticle magnetic moments and their
stability with respect to reorientations is a problem of prominent theoretical
and practical importance. In fact, it is related to the stochastic and
nonlinear dynamics and to the thermal stability of the magnetic moments in
nanoparticle devices including magnetic storage ones. \cite{PEW,DFT} At low
temperatures, when thermal fluctuations are negligible, the main interest is in
the dynamics and stability in time-dependent external magnetic fields. In this
case, the problem is usually reduced to the search for solutions of the
deterministic Landau-Lifshitz equation\cite{LL} and to the analysis of their
stability. These investigations are also strongly motivated by the possibility
of fast switching of the nanoparticle magnetic
moments.\cite{BWH,ABB,BFH,KR,SCC,SW}

Due to thermal fluctuations, the dynamics of the nanoparticle magnetic moments
becomes stochastic and nonzero probabilities of their transition from one
stable state to another appear. In this case, the dynamics can be described by
the Fokker-Planck equation that corresponds to the stochastic Landau-Lifshitz
equation.\cite{B} At present this approach is widely used for studying magnetic
properties of nanoparticle systems at finite temperatures, including magnetic
relaxation.\cite{B,KG,CCK,G,GL,DT,DLT}

In this paper, we use the deterministic and the stochastic Landau-Lifshitz
equations to study some effects induced by the rotating magnetic field in
systems of purely deterministic and weakly superparamagnetic nanoparticles.
More precisely, we are interested in the effects that arise from the different
dynamical states of the up and down magnetic moments. These states are
generated by the magnetic field rotating in the plane perpendicular to the
up-down axis and they are different even if the static magnetic field along
this axis is absent. The reason is that the magnetic moments have a
well-defined direction (counterclockwise) of the natural precession, and so the
rotating field effectively interacts only with the up \textit{or} down magnetic
moments. We note in this context that some properties of the solutions of the
deterministic Landau-Lifshitz equation were previously considered in the
context of ferromagnetic resonance,\cite{W} nonlinear magnetization
dynamics,\cite{TJS,BSM} and switching of magnetization in cylinders \cite{K}
and spherical nanoparticles.\cite{MBSM} But, to the best of our knowledge, the
above mentioned effects have not been investigated before.

The paper is organized as follows. In Sec.~II, we describe the model and the
underlying assumptions. In Sec.~III, we reduce the deterministic
Landau-Lifshitz equation to the algebraic equations that describe the
steady-state forced precession of the nanoparticle magnetic moments and derive
a criterion of its stability. In the same section, we apply the
results for studying the small-angle precession and switching of the magnetic
moments. The effects in nanoparticle systems that arise from the simultaneous
action of the thermal fluctuations and rotating field are considered in
Sec.~IV. Here we calculate the mean residence times that the driven magnetic
moments reside in the up and down states, and apply these results to study the
induced magnetization and magnetic relaxation in nanoparticle systems. We
summarize our findings in Sec.~V. Finally, in the Appendix we specify the used
coordinate systems.

\section{DESCRIPTION OF THE MODEL}

We consider a uniaxial ferromagnetic nanoparticle with spatially uniform
magnetization which is characterized by the anisotropy field $H_{a}$ and the
magnetic moment $\mathbf{m} = \mathbf{m}(t)$ of fixed length $|\mathbf{m}| =
m$. The assumption of uniform magnetization is valid for uniform nanoparticles
if the exchange length, i.e., the length scale below which the exchange
interaction is predominant (for typical magnetic recording materials its order
of magnitude is 5--10 nm), exceeds the nanoparticle size. In other cases, e.g.,
for coated nanoparticles, it can be considered as a first approximation. We
assume also that the static magnetic field $\mathbf{H}$ is applied along the
easy axis of magnetization (the $z$ axis), and the circularly polarized
magnetic field $\mathbf{h(t)}$ is applied in the $xy$ plane (see Fig.~1), i.e.,
$\mathbf{H} = H\mathbf{e}_{z}$ and
\begin{equation}
    \mathbf{h}(t) = h\cos(\omega t)\mathbf{e}_{x} +
    \rho h\sin(\omega t)\mathbf{e}_{y}.
    \label{eq:def_h}
\end{equation}
Here $\mathbf{e}_{x}$, $\mathbf{e}_{y}$, and $\mathbf{e}_{z}$ are the unit
vectors along the corresponding axes of the Cartesian coordinate system $xyz$,
$h = |\mathbf{h}(t)|$, $\omega$ is the frequency of rotation of
$\mathbf{h}(t)$, and $\rho=-1$ or $+1$ that corresponds to the clockwise or
counterclockwise rotation of $\mathbf{h}(t)$, respectively. We write the
magnetic energy of such a nanoparticle as
\begin{equation}
    W = -\frac{H_{a}}{2m}m_{z}^{2} - Hm_{z} - \mathbf{m}\cdot
    \mathbf{h}(t),
    \label{eq:def_W}
\end{equation}
where $m_{z} = \mathbf{m} \cdot \mathbf{e}_{z}$ is the $z$ component of
$\mathbf{m}$, and the dot denotes the scalar product.

In the deterministic case, we describe the dynamics of the nanoparticle
magnetic moment by the Landau-Lifshitz equation\cite{LL}
\begin{equation}
    \dot\mathbf{m} = -\gamma\mathbf{m}\times\mathbf{H}_{eff}
    -\frac{\lambda\gamma}{m}\,\mathbf{m}\times
    (\mathbf{m}\times\mathbf{H}_{eff}).
    \label{eq:L-L}
\end{equation}
Here $\gamma(>0)$ is the gyromagnetic ratio, $\lambda(>0)$ is the dimensionless
damping parameter, the cross denotes the vector product, and
\begin{equation}
    \mathbf{H}_{eff} = -\frac{\partial W}{\partial \mathbf{m}} =
    \mathbf{h}(t) + \bigg(H_{a}\frac{m_{z}}{m} + H\bigg)
    \mathbf{e}_{z}
    \label{eq:def_H_ef}
\end{equation}
is the effective magnetic field acting on $\mathbf{m}$.

If the magnetic moment interacts with a heat bath, we use the stochastic
Landau-Lifshitz equation\cite{B}
\begin{equation}
    \dot\mathbf{m} = -\gamma\mathbf{m}\times(\mathbf{H}_{eff} +
    \mathbf{n}) - \frac{\lambda\gamma}{m}\,\mathbf{m}\times
    (\mathbf{m}\times\mathbf{H}_{eff}),
    \label{eq:st_L-L}
\end{equation}
where $\mathbf{n} = \mathbf{n}(t)$ is the thermal magnetic field with zero mean
and correlation functions $\langle n_{\alpha} (t_1) n_{\beta}(t_2) \rangle =
2\Delta\delta_{\alpha\beta}\delta(t_2 - t_1)$, $n_{\alpha}(t)$ ($\alpha =
x,y,z$) are the Cartesian components of $\mathbf{n} (t)$, $\Delta = \lambda
k_{B}T/\gamma m$ is the intensity of the thermal field, $k_{B}$ is the
Boltzmann constant, $T$ is the absolute temperature, $\delta_{\alpha\beta}$ is
the Kronecker symbol, $\delta(t)$ is the Dirac $\delta$ function, and the
angular brackets denote averaging with respect to the sample paths of
$\mathbf{n}(t)$. According to Eq.~(\ref{eq:st_L-L}), the conditional
probability density $P = P(\theta,\varphi,t| \theta',\varphi', t')$ ($t \geq
t'$) that describes the statistical properties of $\mathbf{m}$ in the terms of
the polar $\theta$ and azimuthal $\varphi$ angles, satisfies the (forward)
Fokker-Planck equation\cite{B,DY}
\begin{eqnarray}
    \frac{\partial}{\partial t}P\! &=& \! -\frac{\partial}{\partial\theta}
    (f_{1} + \gamma^{2}\Delta\cot\theta)P + \gamma^{2}\Delta\frac
    {\partial^{2}}{\partial\theta^{2}}P\nonumber\\[3pt]
    &&\! -\,\frac{\partial}{\partial\varphi}f_{2}P + \frac{\gamma^{2}
    \Delta}{\sin^{2}\theta}\frac{\partial^{2}}{\partial\varphi^{2}}P
    \label{eq:fw_F-P}
\end{eqnarray}
with
\begin{equation}
    \begin{array}{c}
    f_{1} = -\displaystyle\frac{\gamma}{m\sin\theta}\left(\lambda\sin\theta
    \frac{\partial}{\partial\theta} + \frac{\partial}{\partial\varphi}
    \right)W, \\[16pt]
    f_{2} = \displaystyle\frac{\gamma}{m\sin^{2}\theta}
    \left(\sin\theta\frac{\partial}{\partial\theta} -
    \lambda\frac{\partial}{\partial\varphi}\right)W.
    \end{array}
    \label{eq:f1_f2}
\end{equation}

\section{DYNAMICAL EFFECTS}

\subsection{Equations for the forced precession}

To study the forced precession of the nanoparticle magnetic moment and its
stability with respect to small perturbations, we use the Landau-Lifshitz
equation (\ref{eq:L-L}) and represent $\mathbf{m}(t)$ in the form
\begin{equation}
    \mathbf{m}(t) = \mathbf{m}_{0}(t) + \mathbf{m}_{1}(t),
    \label{eq:rep_m}
\end{equation}
where $\mathbf{m}_{0}(t)$ describes the steady-state precession of
$\mathbf{m}(t)$, and $\mathbf{m}_{1}(t)$ is a small deviation from
$\mathbf{m}_{0}(t)$. Since $|\mathbf{m}(t)| = |\mathbf{m}_{0}(t)| = m$, it is
convenient to introduce the unit vector $\mathbf{u} = \mathbf{m}_{0}(t)/m$ and
a small dimensionless vector $\mathbf{v} = \mathbf{m}_{1}(t)/m$ ($ v =
|\mathbf{v}| \ll 1$). According to this, we decompose the effective magnetic
field (\ref{eq:def_H_ef}) into the zeroth-order (in $v$) vector
\begin{equation}
    \mathbf{H}_{eff}^{(0)} = \mathbf{h}(t) + (H_{a} u_{z} + H)
    \mathbf{e}_{z}
    \label{eq:def_H_ef(0)}
\end{equation}
and the first-order one
\begin{equation}
    \mathbf{H}_{eff}^{(1)} = H_{a} v_{z}\mathbf{e}_{z}.
    \label{eq:def_H_ef(1)}
\end{equation}
Substituting Eq.~(\ref{eq:rep_m}) and the effective field $\mathbf{H}_{eff} =
\mathbf{H}_{eff}^{(0)} + \mathbf{H}_{eff}^{(1)}$ into Eq.~(\ref{eq:L-L}) and
keeping the terms of the zeroth order, we end up with the following equation for
$\mathbf{u}$:
\begin{equation}
    \dot{\mathbf{u}} = -\gamma\mathbf{u}\times
    \mathbf{H}_{eff}^{(0)} - \lambda\gamma\mathbf{u}\times
    (\mathbf{u}\times\mathbf{H}_{eff}^{(0)}).
    \label{eq:L-L_steady}
\end{equation}

Introducing, as usual, the rotating Cartesian coordinate system $x'y'z'$ (see
Fig.~1 and the Appendix) and assuming that in this coordinate system the components
$u_{x'}$, $u_{y'}$, and $u_{z'} = u_{z}$ of the vector $\mathbf{u}$ do not
depend on time, Eq.~(\ref{eq:L-L_steady}) can be reduced to a system of
algebraic equations. Indeed, using the relations
\begin{equation}
    \begin{array}{c}
    \mathbf{u} =  u_{x'}\mathbf{e}_{x'} +  u_{y'}\mathbf{e}_{y'}
    +  u_{z}\mathbf{e}_{z},
    \\[10pt]
    \dot{\mathbf{u}} = \rho\omega(- u_{y'}\mathbf{e}_{x'} +
     u_{x'}\mathbf{e}_{y'}),
    \\[10pt]
    \mathbf{H}_{eff}^{(0)} = h\mathbf{e}_{x'} + (H_{a} u_{z}
    + H)\mathbf{e}_{z}
    \end{array}
    \label{eq:rel3}
\end{equation}
that follow from Eqs.~(\ref{eq:def_x'y'})--(\ref{eq:rel1}) and taking the $x'$,
$y'$, and $z$ components of Eq.~(\ref{eq:L-L_steady}), we obtain
\begin{equation}
    \begin{array}{c}
    \lambda u_{x'}(u_{z}^2 + \tilde{H} u_{z} + \tilde{h} u_{x'}) +
    u_{y'}(u_{z} + \tilde{H} - \rho\tilde{\omega}) = \lambda \tilde{h},
    \\[10pt]
    u_{x'}(u_{z} + \tilde{H} - \rho\tilde{\omega}) - \lambda u_{y'}
    (u_{z}^2 + \tilde{H} u_{z} + \tilde{h} u_{x'}) = \tilde{h} u_{z},
    \\[10pt]
    \lambda u_{z}(u_{z}^2 + \tilde{H} u_{z} + \tilde{h} u_{x'}) -
    \lambda (u_{z} + \tilde{H}) = \tilde{h} u_{y'},
    \end{array}
    \label{eq:L-L_comp}
\end{equation}
where $\tilde{H} = H/H_{a}$, $\tilde{h} = h/H_{a}$, $\tilde{\omega} = \omega/
\omega_{r}$, and $\omega_{r} = \gamma H_{a}$. A simple analysis of this system
shows that $u_{x'}$ and $u_{y'}$ are readily expressed through $u_{z}$:
\begin{equation}
    u_{x'} = \displaystyle\frac{1 -  u_{z}^2}{\tilde{h} u_{z}}(u_{z} +
    \tilde{H} - \rho\kappa), \quad
    u_{y'} = -\rho\frac{\lambda\kappa}{\tilde{h}}(1 -  u_{z}^2)\,
    \label{eq:eq uxuy}
\end{equation}
with $\kappa = \tilde{\omega} / (1+\lambda^2)$, and $u_{z}$ satisfies the
equation
\begin{equation}
    \tilde{h}^2 = \frac{1 -  u_{z}^2}{ u_{z}^2}[(u_{z} + \tilde{H} -
    \rho\kappa)^{2} + (\lambda\kappa u_{z})^{2}].
    \label{eq:eq uz}
\end{equation}
It is not difficult to verify that Eqs.~(\ref{eq:eq uxuy}) and (\ref{eq:eq uz})
preserve the condition $\mathbf{u}^{2} = 1$. Note also that the components
$u_{x}$ and $u_{y}$ of $\mathbf{u}$ in the initial coordinate system $xyz$ are
expressed in terms of $u_{x'}$ and $u_{y'}$ as follows:
\begin{equation}
    \begin{array}{c}
     u_{x} =  u_{x'}\cos(\omega t) - \rho u_{y'}\sin(\omega t),
    \\[10pt]
     u_{y} = \rho u_{x'}\sin(\omega t) +  u_{y'}\cos(\omega t).
    \end{array}
    \label{eq:rel7}
\end{equation}

\subsection{Stability criterion}

Next, assuming that the solution of Eq.~(\ref{eq:eq uz}) is known, we derive a
stability criterion for the steady-state precession of $\mathbf{m}$. To this
end, using Eqs.~(\ref{eq:L-L}), (\ref{eq:L-L_steady}) and (\ref{eq:rep_m}), we
write the linear differential equation{\setlength\arraycolsep{2pt}
\begin{eqnarray}
    \dot{\mathbf{v}} &=& -\, \gamma\mathbf{v}\times\mathbf{H}_{eff}
    ^{(0)} - \gamma\mathbf{u}\times\mathbf{H}_{eff}^{(1)} -
    \lambda\gamma[\mathbf{v}(\mathbf{u}\cdot\mathbf{H}_{eff}^{(0)})
    \nonumber\\[4pt]
    && +\,\mathbf{u}(\mathbf{v}\cdot\mathbf{H}_{eff}^{(0)})
    + \mathbf{u}\times(\mathbf{u}\times\mathbf{H}_{eff}^{(1)})]
    \label{eq:perturb1}
\end{eqnarray}}that describes the evolution of small deviations $\mathbf{v}$.
Since Eq.~(\ref{eq:L-L}) conserves $|\mathbf{m}|$, the condition $2\mathbf{u}
\cdot \mathbf{v} +  v^{2} = 0$ always holds. This means that, with linear
accuracy in $ v$, the vector $\mathbf{v}$ is perpendicular to $\mathbf{u}$ for
all $t$. Therefore, it is convenient to introduce the rotating Cartesian
coordinate system $x''y''z''$ (see Fig.~1 and Appendix) in which the vectors
$\mathbf{u}$ and $\mathbf{v}$ are represented as
\begin{equation}
    \mathbf{u} = \mathbf{e}_{z''}, \quad \mathbf{v} =  v_{x''}(t)
    \mathbf{e}_{x''} +  v_{y''}(t)\mathbf{e}_{y''},
    \label{eq:mu_xi}
\end{equation}
and so the condition $\mathbf{u}\cdot\mathbf{v} = 0$ holds automatically.

In this coordinate system, the $z''$ component of Eq.~(\ref{eq:perturb1}) is
satisfied identically because according to the condition $d(\mathbf{u} \cdot
\mathbf{v})/dt = 0$ and Eq.~(\ref{eq:L-L_steady}) the relation
\begin{equation}
    \big[\dot{\mathbf{v}} + \gamma\mathbf{v}\times
    \mathbf{H}_{eff}^{(0)} + \lambda\gamma\mathbf{u}(\mathbf{v}\cdot
    \mathbf{H}_{eff}^{(0)})\big]\cdot\mathbf{e}_{z''} = 0
    \label{eq:rel5}
\end{equation}
always takes place. Projecting Eq.~(\ref{eq:perturb1}) onto the $x''$ and $y''$
axes and using Eqs.~(\ref{eq:def_x"y"z"})--(\ref{eq:rel2}), as well as  the results of
the previous section, we obtain after straightforward calculations
\begin{equation}
    \begin{array}{c}
    \dot v_{x''} = -\lambda\omega_{1} v_{x''} -
    \omega_{2} v_{y''},
    \\[10pt]
    \dot v_{y''} = \omega_{3} v_{x''} -
    \lambda\omega_{4} v_{y''},
    \end{array}
    \label{eq:perturb2}
\end{equation}
where $\omega_{n} = \omega_{r}\tilde{\omega}_{n}$ and
\begin{equation}
    \begin{array}{c}
    \tilde{\omega}_{1} = u_{z}^{2} + \displaystyle\frac{1}{u_{z}}
    [\tilde{H} - \rho\kappa(1 - u_{z}^{2})], \\[10pt]
    \tilde{\omega}_{2} = 1 + \displaystyle\frac{1}{u_{z}}
    [\tilde{H} - \rho\kappa(1 + \lambda^{2}u_{z}^{2})], \\[10pt]
    \tilde{\omega}_{3} = u_{z}^{2} + \displaystyle\frac{1}{u_{z}}
    [\tilde{H} - \rho\kappa(1 + \lambda^{2}u_{z}^{2})], \\[10pt]
    \tilde{\omega}_{4} = 1 + \displaystyle\frac{1}{u_{z}}
    [\tilde{H} - \rho\kappa(1 - u_{z}^{2})].
    \end{array}
    \label{eq:def_omega}
\end{equation}

Thus, in the first-order approximation, the stability of the steady-state
precession of the nanoparticle magnetic moment is defined by the stability of
the stationary solution $v_{x''} = v_{y''} = 0$, or the fixed point $(0,0)$, of
the system (\ref{eq:perturb2}). A complete solution of the last problem is well
known (see, for example in Ref.~[\onlinecite{P}]), and is based on the analysis
of the roots
\begin{equation}
    \delta_{\pm} = -\frac{\lambda}{2}(\omega_{1} + \omega_{4})
    \pm\frac{1}{2}\sqrt{\lambda^{2}(\omega_{4} - \omega_{1})^{2}
    - 4\omega_{2}\omega_{3}}
    \label{eq:def_delta}
\end{equation}
of the characteristic equation $(\delta + \lambda\omega_{1})(\delta +
\lambda\omega_{4}) + \omega_{2}\omega_{3} = 0$ corresponding to this system. In
particular, a criterion of the asymptotic stability of the forced precession
has the form $\textrm{Re}\, \delta_{+} < 0$ or
\begin{equation}
    \lambda(\tilde{\omega}_{1} + \tilde{\omega}_{4}) >
    \textrm{Re}\sqrt{\lambda^{2}(\tilde{\omega}_{4} - \tilde{\omega}_{1})^{2}
    - 4\tilde{\omega}_{2}\tilde{\omega}_{3}}\,.
    \label{eq:stab_criterion}
\end{equation}

In the following we apply the above general results to study the precessional
dynamics in the cases of small precession angles and zero static magnetic
field.

\subsection{Small precession angles}

In this case, we assume that the precession angles $\theta_{\sigma}$ (see
Fig.~2) of the magnetic moments with $u_{z} > 0$ ($\sigma = +1$) and $u_{z} <
0$ ($\sigma = -1$) are small, i.e., $\theta_{\sigma} \ll 1$. Then $u_{z}$ can
be represented in the form $u_{z} = \sigma(1-\epsilon^{2}/2)$, where according
to Eq.~(\ref{eq:eq uz}) a small parameter $\epsilon^{2}$ is given by
\begin{equation}
    \epsilon^{2} = \displaystyle\frac{\tilde{h}^{2}}
    {(\sigma + \tilde{H} - \rho\kappa)^{2} + \lambda^{2}\kappa^{2}},
    \label{eq:epsilon}
\end{equation}
and so Eq.~(\ref{eq:eq uxuy}) yields with linear accuracy in $\epsilon$
\begin{equation}
    \begin{array}{c}
    u_{x'} = \tilde{h}\displaystyle\frac{1 + \sigma\tilde{H} -
    \sigma\rho\kappa}{(\sigma + \tilde{H} - \rho\kappa)^{2} +
    \lambda^{2}\kappa^{2}},
    \\[14pt]
    u_{y'} = -\tilde{h}\displaystyle\frac{\rho\lambda\kappa}
    {(\sigma + \tilde{H} - \rho\kappa)^{2} + \lambda^{2}\kappa^{2}}.
    \end{array}
    \label{eq:mu_x_y}
\end{equation}
We emphasize that, even though the static magnetic field is absent, i.e.,
$\tilde{H} = 0$, the dynamics of the up ($\sigma = +1$) and down ($\sigma =
-1$) magnetic moments is quite different. The reason is that the natural
precession of the magnetic moments is counterclockwise, and so only up or down
magnetic moments have the direction of the natural precession that coincides
with the direction of the magnetic field rotation. In other words, the magnetic
field \textit{rotating} in the plane perpendicular to the easy axis of
magnetization breaks the degeneracy between the up and down states of the
magnetic moment.

Our analysis shows that the small-angle precession is stable only if
$\tilde{\omega}_{1} + \tilde{\omega}_{4} > 0$. Writing $\tilde{\omega}_{1} +
\tilde{\omega}_{4}$ with quadratic accuracy in $\epsilon$,
\begin{equation}
    \tilde{\omega}_{1} + \tilde{\omega}_{4} = 2 + 2\sigma\tilde{H}
    - \epsilon^{2}(1 - \sigma\tilde{H} + 2\sigma\rho\kappa),
    \label{eq:summ}
\end{equation}
and solving the equation $\tilde{\omega}_{1} + \tilde{\omega}_{4} = 0$ with
respect to $\tilde{H}$, we find the critical magnetic field
\begin{equation}
    \tilde{H}_{cr} = -\sigma + \tilde{h}^{2}\frac{\sigma + \rho\kappa}
    {(1 + \lambda^{2})\kappa^{2}}
    \label{eq:switch}
\end{equation}
that separates the stable and unstable precession for a given state $\sigma$.
The steady precession is stable either for $\tilde{H} > \tilde{H}_{cr}|_{\sigma
= +1}$ or for $\tilde{H} < \tilde{H}_{cr}|_{\sigma = -1}$. Because at $\tilde{H}
= \tilde{H}_{cr}$ the precession in the state $-\sigma$ is stable, the
switching of the nanoparticle magnetic moments from the unstable state $\sigma$
to the stable state $-\sigma$ occurs. If $\rho = \sigma$, then the rotating
field always decreases the stability of the precession, i.e., $\tilde{H}_{cr}
|_{\sigma = +1} > -1$ and $\tilde{H}_{cr} |_{\sigma = -1} < +1$. On the
contrary, if $\rho = -\sigma$ then depending on the reduced frequency $\kappa$
the rotating field can both decrease (if $\kappa < 1$) and increase (if $\kappa
> 1$) the stability. The largest stabilization effect is achieved at $\kappa =
2$. Note also that since $\epsilon \ll 1$ and usually $\lambda < 1$, the
formula (\ref{eq:switch}) is valid only if the condition $\kappa \gg \tilde{h}$
holds.

As an important illustrative example, we consider the nanoparticle system with
the same number $N/2$ of the up and down states. In this case the dynamical
(dimensionless) magnetization of the system $\mu_{d} = (1/N)\sum_{i=1}^
{N}u_{zi}$ ($i$ labels the nanoparticles) takes the form $\mu_{d} = (1/2)\sum_
{\sigma}\sigma \cos\theta_{\sigma}$ or, since $\theta_{\sigma} \ll 1$, $\mu_{d}
= (\theta_{-1}^{2} - \theta_{+1}^{2})/4$. Assuming that $\tilde{H} = 0$ and
using the formula
\begin{equation}
    \theta_{\sigma} = \frac{\tilde{h}}{\sqrt{(1 - \sigma\rho\kappa)^{2} +
    \lambda^{2}\kappa^{2}}}
    \label{eq:theta_sigma}
\end{equation}
that follows from the relation $\sin \theta_{\sigma} = \epsilon$ and
Eq.~(\ref{eq:epsilon}), this quantity can be written in the form
\begin{equation}
    \mu_{d} = -\tilde{h}^{2}\frac{\rho\kappa}{[1 + (1 + \lambda^{2})
    \kappa^{2}]^{2} - 4\kappa^{2}}.
    \label{eq:mu_d}
\end{equation}

This result shows that (i) the magnetization $\mu_{d}$ is a purely dynamical
effect, i.e., $\mu_{d}$ = 0 if $\kappa = \tilde{\omega} / (1+\lambda^2)= 0$,
(ii) the direction of magnetization and the direction of magnetic field
rotation follow the left-hand rule, and (iii) the dependence of $\mu_{d}$ on
$\kappa$ always exhibits a resonant character. The maximum of $\mu_{d}$ occurs
at $\kappa = \kappa_{m}$, where
\begin{equation}
    \kappa_{m} = \frac{1}{\sqrt{3}(1 + \lambda^{2})}
    \sqrt{1 - \lambda^{2} + 2\sqrt{1 + \lambda^{2} + \lambda^{4}}},
    \label{eq:kappa m}
\end{equation}
and $\mu_{d}|_{\kappa = \kappa_{m}} = -\rho(\tilde{h}/2\lambda)^{2}$ for
$\lambda \ll 1$. If $\tilde{h} \ll \lambda$ then the dynamical magnetization is
small but, as we will show later, it can be considerably enhanced by thermal
fluctuations.

\subsection{Zero static magnetic field}

In the case of zero static magnetic field, $\tilde{H} = 0$, we rewrite
Eq.~(\ref{eq:eq uz}) in the form $\tilde{h} = F_{\rho}(u_z)$, where
\begin{equation}
    F_{\rho}(x) = \frac{\sqrt{1 -  x^2}}{|x|}\sqrt{(x - \rho\kappa)^{2}
    + (\lambda\kappa x)^{2}}
    \label{eq:eq uz1}
\end{equation}
($-1 \leq x \leq 1$). According to the definition, the function $F_{\rho}(x)$
satisfies the conditions $F_{\rho}(-x) = F_{-\rho}(x)$, $F_{\rho}(-1) =
F_{\rho}(+1) = 0$, $F_{\rho}(x) \to \infty$ as $x \to 0$, and it has a local
minimum at $x = \rho x_{1}\,(x_{1}>0)$ and a local maximum at $x = \rho x_{2}
\,(x_{2}>0)$, see Fig.~3.

A detailed analysis shows that for fixed $\rho$ the precession of the
nanoparticle magnetic moment in the state $\sigma = -\rho$ is stable for all
values of $\tilde{h}$. In other words, the unique solution of the equation
$\tilde{h} = F_{\rho}(u_{z})$ with $\text{sgn}\, u_{z} = -\rho$ always exists
and is stable. In this case, the dependence of $u_{z}$ on $\tilde{h}$ is shown
in Fig.~4, curve 1.

The precession of the magnetic moment in the state $\sigma = \rho$ (when
$\text{sgn}\, u_{z} = \rho$) exhibits qualitatively different behavior
depending on $\tilde{h}$. It is stable only if $|u_{z}| > x_{2}$ that implies
that $\tilde{h} < \tilde{h}_{cr} = F_{\rho}(\rho x_{2})$ (see Fig.~4, curve 2).
At $\tilde{h} = \tilde{h}_{cr} + 0$ the solutions $u_{z}|_{\sigma = \rho} =
\rho x_{2}$ and $u_{z}|_{\sigma = \rho} = \rho x_{3}\,(x_{3} > 0)$ are
unstable, and the magnetic moment switches from the state with $u_{z}|_{\sigma
= \rho} = \rho x_{2}$ to the new state with $u_{z}|_{\sigma = -\rho} = -\rho
x_{4}\,(x_{4} > 0)$. As stated above, the new state is stable for all
$\tilde{h}$, and so the reverse transition does not occur for fixed $\rho$.

In the important case of small driven frequency, $\kappa \ll 1$, that is easily
accessible to experimental investigation, the analysis of the stability of the
forced precession can be done analytically. In particular, we found that
\begin{equation}
    u_{z} = \sigma\sqrt{1-\tilde{h}^{2}} -
    \rho\frac{\tilde{h}^{2}}{1-\tilde{h}^{2}}\kappa
    \label{eq:as1}
\end{equation}
if $1 - \tilde{h}^{2} \gg \kappa^{2/3}$ and $u_{z}|_{\sigma = -\rho} = -\rho
\kappa /\tilde{h}$ if $\tilde{h} \gg 1$. Using the approximate representations
\begin{equation}
    \tilde{\omega}_{1} = \tilde{\omega}_{3} = u_{z}^{2} - \rho\frac{\kappa}
    {u_{z}}, \quad
    \tilde{\omega}_{2} = \tilde{\omega}_{4} = 1 - \rho\frac{\kappa}{u_{z}},
    \label{eq:om}
\end{equation}
we showed explicitly that the stability criterion (\ref{eq:stab_criterion}) for
$\sigma = \rho$ is reduced to $|u_{z}| > x_{2}$ and the critical amplitude of
the rotating field is given by $\tilde{h}_{cr} = 1 - (3/2)\kappa^{2/3}$.
Finally, solving the equations $dF_{\rho}(x)/dx = 0$ and $\tilde{h}_{cr} =
F_{\rho}(x)$ with respect to $x$, we derived the asymptotic formulas $x_{1} =
\kappa$, $x_{2} = \kappa^{1/3}$, $x_{3} = \kappa/2$, and $x_{4} = 2\kappa^
{1/3}$. It is important to note that although the jump $\Delta u_{z} = x_{2} +
x_{4}$ that occurs under switching of $u_{z}$ tends to zero as $\kappa \to 0$,
it can be sizeable even for very small $\kappa$ (for example, $\Delta u_{z} =
0.3$ if $\kappa = 10^{-3}$).

We emphasize also that this truly remarkable phenomenon, the switching of the
nanoparticle magnetic moments under the action of the rotating field, results
from the existence of the natural precession of the magnetic moments. It occurs
only in those nanoparticles for which the condition $\sigma\rho = +1$ holds,
i.e., if the direction of the magnetic field rotation coincides with the
direction of the natural precession of the magnetic moments.

\section{THERMAL EFFECTS}

\subsection{Mean residence times}

If the magnetic moments interact with a heat bath then their dynamics becomes
stochastic and is described by the forward Fokker-Planck equation
(\ref{eq:fw_F-P}). In this case, due to thermal fluctuations, the magnetic
moments can perform random transitions from the one state $\sigma$ to the other
$-\sigma$. Our aim is to study how the rotating magnetic field influences the
mean residence times $t_{\sigma}$ that the magnetic moments dwell in these
states at $\tilde{H} = 0$. In principle, the problem can be solved on the basis
of Eq.~(\ref{eq:fw_F-P}). Specifically, this approach has been used in the case
of ac magnetic field linearly polarized along the easy axis of
magnetization.\cite{PR} However, since the mean residence times can be readily
expressed through the mean-first passage times, for solving this problem it is
convenient to use the backward Fokker-Planck equation\cite{HT,DY}
\begin{eqnarray}
    \frac{\partial}{\partial t'}P\! &=& \! -(f_{1}' + \gamma^{2}\Delta\cot
    \theta')\frac{\partial}{\partial\theta'}P - \gamma^{2}\Delta
    \frac{\partial^{2}}{\partial\theta'^{2}}P\nonumber\\[3pt]
    &&\! -\,f_{2}'\frac{\partial}{\partial\varphi'}P - \frac{\gamma^{2}
    \Delta}{\sin^{2}\theta'}\frac{\partial^{2}}{\partial\varphi'^{2}}P
    \label{eq:bw_F-P}
\end{eqnarray}
[$f_{1,2}' = f_{1,2}(\theta',\varphi',t')$], which is equivalent to
Eq.~(\ref{eq:fw_F-P}). Within this framework, we are able to calculate
$t_{\sigma}$ in some particular cases. But because of the procedure is rather
complicated from the mathematical point of view (details will be published
elsewhere), here we use a crude approximation that leads, however, to
qualitatively the same results.

In the considered case of small-angle precession, the magnetic moments of
weakly superparamagnetic particles (when $a = H_{a}m/2k_{B}T \gg 1$) spend
almost all time near the conic surfaces with the cone angles
(\ref{eq:theta_sigma}). We assume that if these imaginary surfaces are replaced
by the reflecting surfaces, then the rotating field terms can be eliminated
from Eq.~(\ref{eq:bw_F-P}). Then, replacing also the conditional probability
density $P$ by $\bar{P} = \bar{P}(\theta,t|\theta', t')$ and taking into
account that $f_{1}' = - (\lambda/2) \omega_{r}\sin 2\theta'$,
Eq.~(\ref{eq:bw_F-P}) reduces to the simpler form
\begin{equation}
    \frac{\partial}{\partial t'}\bar{P} = \bigg(\frac{\lambda}{2}\,\omega_{r}
    \sin 2\theta' - \gamma^{2}\Delta\cot\theta'\bigg)\frac{\partial}{\partial
    \theta'}\bar{P} - \gamma^{2}\Delta\frac{\partial^{2}}{\partial\theta'^{2}}
    \bar{P}.
    \label{eq:F-P}
\end{equation}

Next we use a standard procedure \cite{HTB} to define the mean first passage
times for the magnetic moments in the states $\sigma$
\begin{equation}
    T_{\sigma}(\theta') = \int_{0}^{\infty} dt \int_{
    \theta_{\sigma}^{(1)}}^{\theta_{\sigma}^{(2)}} d\theta
    P(\theta,t|\theta',0)
    \label{eq:def_T}
\end{equation}
and to derive from Eq.~(\ref{eq:F-P}) the ordinary differential equation for
these quantities
\begin{equation}
    \frac{d^{2}T_{\sigma}(\theta')}{d\theta'^{2}} +
    (\cot\theta' - a\sin 2\theta')\frac{dT_{\sigma}
    (\theta')}{d\theta'} + at_{r} =0.
    \label{eq:eq_T}
\end{equation}
Here, $\theta' \in (\theta_{+1},\pi/2)$ if $\sigma = +1$, $\theta' \in
(\pi/2,\pi - \theta_{-1})$ if $\sigma = -1$, $\theta_{+1}^{(1)} = \theta_{+1}$,
$\theta_{+1}^{(2)} = \theta_{-1}^{(1)} = \pi/2$, $\theta_{-1} ^{(2)} = \pi -
\theta_{-1}$, and $t_{r} = 2/\lambda \omega_{r}$ is the characteristic
relaxation time of the precessional motion of the magnetic moment. Solving
Eq.~(\ref{eq:eq_T}) with the absorbing and reflecting boundary conditions,
i.e., $T_{\sigma}(\pi/2) = 0$ and $dT_{\sigma}(\theta') /d\theta'| _{\theta' =
\pi(1 - \sigma)/2 + \sigma \theta_{\sigma}} = 0$, respectively, and taking into
account that the desired times $t_{\sigma}$ are readily expressed through
$T_{\sigma} (\theta')$, $t_{\sigma} = 2T_{\sigma} \textbf{(}\pi[1 - \sigma]/2 +
\sigma \theta_ {\sigma} \textbf{)}$, we obtain for $a \gg 1$ and
$\theta_{\sigma} \ll 1$
\begin{equation}
    t_{\sigma} = \frac{t_{r}}{2}\sqrt{\frac{\pi}{a}}\,
    \exp[a(1+\sigma2\tilde{H}_{\sigma})],
    \label{eq:t_sigma2}
\end{equation}
where $\tilde{H}_{\sigma} = -\sigma\tilde{\theta}_{\sigma}^{2}/2$ can by
interpreted as an effective magnetic field acting on the nanoparticles in the
state $\sigma$.

According to this result, the rotating magnetic field decreases the mean
residence times. However, due to the natural precession, the decrease of the
mean times is different for the up and down magnetic moments. As it will be
shown below, this fact causes a strong enhancement of the dynamical
magnetization and leads to a modification of the relaxation law.

\subsection{Induced magnetization}

We define the steady state magnetization of the nano\-particle system in the
rotating magnetic field as $\mu = (1/N)\sum_{i=1}^{N}m_{zi}/m$ ($N \to
\infty$). Denoting the average number of the magnetic moments in the state
$\sigma$ as $N_{\sigma}$ and introducing the probability $p_{\sigma} =
N_{\sigma}/N$ ($p_{+1} + p_{-1} = 1$) that the magnetic moment resides in this
state, we rewrite $\mu$ in the form
\begin{equation}
    \mu = \sum_{\sigma}p_{\sigma}\langle \cos\theta \rangle|_{\sigma},
    \label{eq:mu}
\end{equation}
where $\langle \cos\theta \rangle|_{\sigma} = (1/N_{\sigma}) \sum_{i \in
N_{\sigma}}\cos\theta_{i}$ is the average value of $m_{zi}/m$ in the state
$\sigma$. If $a \gg 1$ then thermal fluctuations are small and $\langle
\cos\theta \rangle|_{\sigma}$ in Eq.~(\ref{eq:mu}) can be replaced by $\sigma
\cos\theta_{\sigma} \approx \sigma(1 - \theta_{\sigma}^{2}/2)$, yielding $\mu =
\mu_{t} + \mu_{td} + \mu_{d}$. Here $\mu_{t} = \sum_{\sigma}\sigma p_{\sigma}$
and $\mu_{td} = \sum_{\sigma}\sigma(2p_{\sigma} - 1)\theta_{\sigma}^{2}/4$ are
the contributions of thermal fluctuations to the total magnetization $\mu$, and
$\mu_{d}$ is the purely dynamical magnetization given by Eq.~(\ref{eq:mu_d}).
Since $\theta_{\sigma} \ll 1$, the condition $\mu_{t}/\mu_{td} \gg 1$ holds,
and so $\mu \approx \mu_{t} + \mu_{d}$. We emphasize that  a decrease of the
temperature (increasing of $a$) decreases the fluctuations of the magnetic
moments, but not the difference between $p_{+1}$ and $p_{-1}$, i.e., $\mu_{t}$.
Moreover, one expects that, similar to the two-level models, $|\mu_{t}|$ grows
with $a$.

Next, taking into account that in the steady state $p_{\sigma} = t_{\sigma}/
(t_{+1} + t_{-1})$ and using Eq.~(\ref{eq:t_sigma2}), we obtain
\begin{equation}
    \mu_{t} = \tanh [a(\tilde{H}_{+1} + \tilde{H}_{-1})].
    \label{eq:ind_magn}
\end{equation}
Comparing $\mu_{t}$ with the magnetization of an Ising paramagnet, $\tanh
(2a\tilde{H})$, we see that the circularly polarized magnetic field induces the
same magnetization $\mu_{t}$ of the nanoparticle system as the external
magnetic field $(\tilde{H}_{+1} + \tilde{H}_{-1})/2$ applied perpendicular to
the polarization plane. It is interesting to note that $(\tilde{H}_{+1} +
\tilde{H}_{-1})/2 = \mu_{d}$, and hence $\mu_{t} = \tanh(2a\mu_{d})$. Since $a
\gg 1$ and $|\mu_{d}| \ll 1$, this relation shows that $\mu_{t}/\mu_{d} \gg 1$
and so $\mu \approx \mu_{t}$, i.e., thermal fluctuations strongly enhance the
dynamical magnetization. In particular, if $a|\mu_{d}| \ll 1$ then
$\mu_{t}/\mu_{d} = 2a$. Like $\mu_{d}$ the dependence of $\mu_{t}$ on $\kappa$
has a resonant character and, as expected, $|\mu_{t}|$ increases with
decreasing temperature (see Fig.~5).

\subsection{Relaxation law}

As a second example, let us consider the thermally activated magnetic
relaxation in the nanoparticle system driven by the rotating field. Since the
transition rate of the nanoparticle magnetic moment from the state $\sigma$ to
the state $-\sigma$ equals $1/t_{\sigma}$, the differential equation that
defines the time-dependent magnetization $\mu(t)$ of this system can be written
in the form
\begin{equation}
    \dot{\mu}(t) = -\mu(t)\left(\frac{1}{t_{+1}} + \frac{1}{t_{-1}}\right)
    - \frac{1}{t_{+1}} + \frac{1}{t_{-1}}.
    \label{eq:eq_law}
\end{equation}
Assuming that $\mu(0)=1$ (we neglect the dynamical magnetization), from
Eq.~(\ref{eq:eq_law}) we obtain the relaxation law
\begin{equation}
    \mu(t) = (1-\mu_{t})\exp(-t/\tau) + \mu_{t},
    \label{eq:rel_law}
\end{equation}
where
\begin{equation}
    \tau = \tau_{0} \frac{\exp[a(\tilde{H}_{+1} - \tilde{H}_{-1})]}
    {\cosh[a(\tilde{H}_{+1} + \tilde{H}_{-1})]}
    \label{eq:rel_time}
\end{equation}
is the relaxation time in the presence of the rotating magnetic field, and
$\tau_{0} = (t_{r}/4)\sqrt{\pi/a}\exp a$ is the relaxation time if the rotating
field is absent.

Thus, the rotating magnetic field decreases the relaxation time ($\tau/
\tau_{0} < 1$) and leads to nonzero magnetization in the long-time limit
[$\mu(\infty) = \mu_{t}$].

\section{CONCLUSIONS}

We have investigated a number of dynamical and thermal effects in nanoparticle
systems that result from the action of a circularly polarized magnetic field
rotating in the plane perpendicular to the easy axes of the nanoparticles. The
main finding is that the dynamics of the nanoparticle magnetic moments, both
deterministic and stochastic, becomes different in the up and down states. It
is important to note that, due to the (counterclockwise) natural precession of
the magnetic moments, the dynamics is different even if the static magnetic
field is absent.

To describe the dynamical effects at zero temperature, we have used the
deterministic Landau-Lifshitz equation. We have solved this equation for
small-angle precession of the magnetic moments and have demonstrated that the
rotating field, depending on its frequency and polarization, can either
decrease or increase the stability of the precession motion. For zero static
field, we have calculated the dynamical magnetization of nanoparticle systems
and predicted the switching effect. This remarkable effect, which consists in
changing the state of the magnetic moments at some critical amplitude of the
rotating magnetic field, occurs only for resonant nanoparticles, i.e., when the
direction of the natural precession of their magnetic moments coincides with
the direction of the magnetic field rotation.

In the case of finite temperatures, we have invoked the backward Fokker-Planck
equation to calculate the mean residence times that the driven magnetic moments
dwell in the up and down states, respectively. On this basis, we have studied
the steady-state magnetization and the features of magnetic relaxation in
systems of weakly superparamagnetic nanoparticles that are driven by the
rotating magnetic field. In particular, we have found that thermal fluctuations
strongly enhance the dynamical magnetization and that the rotating field always
causes a decrease of the relaxation time.

\section*{ACKNOWLEDGMENTS}

S.I.D., T.V.L., and K.N.T. acknowledge the support of the EU through NANOSPIN
contract No NMP4-CT-2004-013545, S.I.D. acknowledges the support of the EU
through a Marie Curie individual fellowship, contract No MIF1-CT-2005-007021,
and P.H. acknowledges the support of the DFG via the SFB 486, project A 6.

\appendix*
\section{ROTATING COORDINATE SYSTEMS}

\subsection{Single-primed coordinate system}

The rotating Cartesian coordinate system $x'y'z'$ is defined by the unit
vectors $\mathbf{e}_{x'}(t)$, $\mathbf{e}_{y'}(t)$, and $\mathbf{e}_{z'}(t)$
that are expressed through the unit vectors of the initial (laboratory)
coordinate system $xyz$ as follows:
\begin{equation}
    \begin{array}{c}
    \mathbf{e}_{x'} = \cos(\omega t)\mathbf{e}_{x} +
    \rho\sin(\omega t)\mathbf{e}_{y},
    \\[12pt]
    \mathbf{e}_{y'} = -\rho\sin(\omega t)\mathbf{e}_{x} +
    \cos(\omega t)\mathbf{e}_{y},
    \end{array}
    \label{eq:def_x'y'}
\end{equation}
$\mathbf{e}_{z'} = \mathbf{e}_{z}$. According to Eqs.~(\ref{eq:def_x'y'}), the
inverse transformation has the form
\begin{equation}
    \begin{array}{c}
    \mathbf{e}_{x} = \cos(\omega t)\mathbf{e}_{x'} -
    \rho\sin(\omega t)\mathbf{e}_{y'},
    \\[12pt]
    \mathbf{e}_{y} = \rho\sin(\omega t)\mathbf{e}_{x'} +
    \cos(\omega t)\mathbf{e}_{y'},
    \end{array}
    \label{eq:inv_x'y'}
\end{equation}
and
\begin{equation}
    \mathbf{\dot{e}}_{x'} = \rho\omega\mathbf{e}_{y'}, \quad
    \mathbf{\dot{e}}_{y'} = -\rho\omega\mathbf{e}_{x'}.
    \label{eq:rel1}
\end{equation}

\subsection{Double-primed coordinate system}

The unit vectors $\mathbf{e}_{x''}(t)$, $\mathbf{e}_{y''}(t)$, and $\mathbf{e}_
{z''}(t)$ of the rotating Cartesian coordinate system $x''y''z''$ are
introduced as
\begin{equation}
    \begin{array}{c}
    \mathbf{e}_{x''} = \displaystyle{\frac{( u_{x'}\mathbf{e}_{x'}
    +  u_{y'}\mathbf{e}_{y'}) u_{z'}}{\sqrt{ u_{x'}^{2} +
     u_{y'}^{2}}} - \sqrt{ u_{x'}^{2} +
     u_{y'}^{2}}\,\mathbf{e}_{z'}},
    \\[22pt]
    \mathbf{e}_{y''} = \displaystyle{\frac{- u_{y'}\mathbf{e}_{x'}
    +  u_{x'}\mathbf{e}_{y'}}{\sqrt{ u_{x'}^{2} +
     u_{y'}^{2}}}},
    \\[22pt]
    \mathbf{e}_{z''} =  u_{x'}\mathbf{e}_{x'} +
     u_{y'}\mathbf{e}_{y'} +  u_{z'}\mathbf{e}_{z'},
    \end{array}
    \label{eq:def_x"y"z"}
\end{equation}
and so the inverse transformation is given by
\begin{equation}
    \begin{array}{c}
    \mathbf{e}_{x'} = \displaystyle{\frac{ u_{x'} u_{z'}
    \mathbf{e}_{x''} -  u_{y'}\mathbf{e}_{y''}}{\sqrt{ u_{x'}^{2}
    +  u_{y'}^{2}}} +  u_{x'}\mathbf{e}_{z''}},
    \\[22pt]
    \mathbf{e}_{y'} = \displaystyle{\frac{ u_{y'} u_{z'}
    \mathbf{e}_{x''} +  u_{x'}\mathbf{e}_{y''}}
    {\sqrt{ u_{x'}^{2} +  u_{y'}^{2}}} +
     u_{y'}\mathbf{e}_{z''}},
    \\[22pt]
    \mathbf{e}_{z'} = -\sqrt{ u_{x'}^{2} +  u_{y'}^{2}}\,
    \mathbf{e}_{x''} +  u_{z'}\mathbf{e}_{z''}.
    \end{array}
    \label{eq:inv_x"y"z"}
\end{equation}
From here and (\ref{eq:rel1}), straightforward calculations yield
\begin{equation}
    \begin{array}{c}
    \mathbf{\dot{e}}_{x''} = \rho\omega u_{z'}\mathbf{e}_{y''},
    \\[10pt]
    \mathbf{\dot{e}}_{y''} = -\rho\omega\left( u_{z'}\mathbf{e}_{x''}
    + \sqrt{ u_{x'}^{2} +  u_{y'}^{2}}\,\mathbf{e}_{z''}\right).
    \end{array}
    \label{eq:rel2}
\end{equation}

\newpage

\begin{figure}
    \centering
    \includegraphics[totalheight=4.5cm]{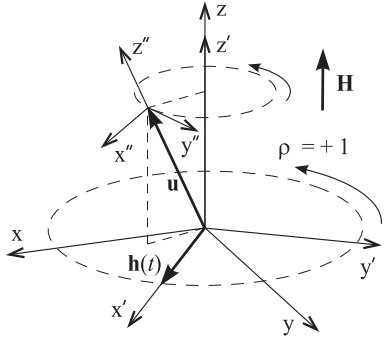}
    \caption{\label{fig1} Schematic representation of the
    model and the used coordinate systems.}
\end{figure}

\begin{figure}
    \centering
    \includegraphics[totalheight=4.5cm]{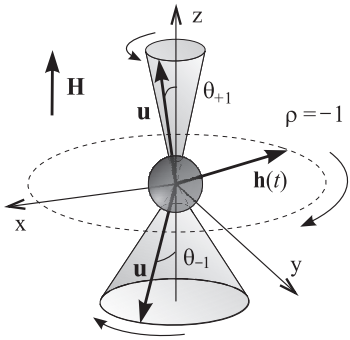}
    \caption{\label{fig2} Sketch of the precession angles for the up
    and down magnetic moments (the arrows depict the directions of their
    natural precession).}
\end{figure}

\begin{figure}
    \centering
    \includegraphics{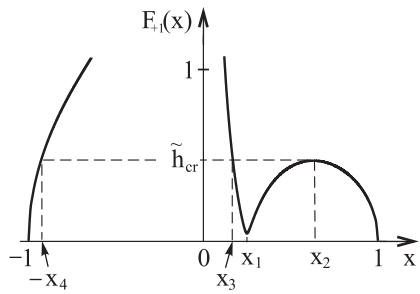}
    \caption{\label{fig3} Plot of the function $F_{+1}(x)$ for
    $\kappa = 0.25$ and $\lambda = 0.2$. If $\kappa \to 0$, then
    $x_{n} \to 0$, $F_{+1}(x_{1}) \to 0$, and $\tilde{h}_{cr} =
    F_{+1}(x_{2}) \to 1$.}
\end{figure}

\begin{figure}
    \centering
    \includegraphics{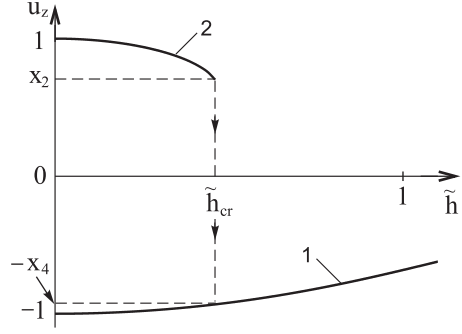}
    \caption{\label{fig4} Plots of the stable solutions of the equation
    $\tilde{h} = F_{+1}(u_z)$ for the same parameters as in Fig.~3. For
    $\rho = -1$ the curves 1 and 2 must be reflected with respect
    to the axis $\tilde{h}$.}
\end{figure}

\begin{figure}
    \centering
    \includegraphics{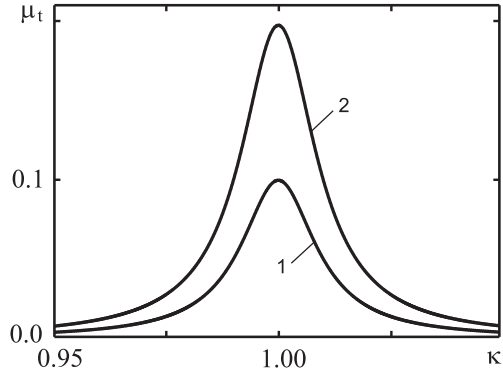}
    \caption{\label{fig5} Plots of the magnetization $\mu_{t}$ vs the
    reduced frequency $\kappa$ for the parameter choice: $\tilde{h} =
    10^{-3}$, $\lambda = 10^{-2}$, $a = 20$ (curve 1) and for $a = 40$
    (curve 2). The temperature in the latter case is two times less than
    in the former one.}
\end{figure}

\end{document}